%% file: main.tex
\documentclass[times, 12pt, a4paper]{article}

\usepackage[utf8]{inputenc}
\usepackage{graphicx}
\usepackage{subfig}
\usepackage{gensymb}
\usepackage{epstopdf}
\usepackage{amsmath}
\usepackage{algorithm}
\usepackage{algorithmic}
\usepackage{authblk}

\newcommand{\specialcell}[2][c]{\begin{tabular}[#1]{@{}c@{}}#2\end{tabular}}
\newcommand*{\rom}[1]{\expandafter\@slowromancap\romannumeral #1@}
\providecommand{\keywords}[1]{\textbf{\textit{Keywords---}} #1}
\newcommand{\locimage}[1]{./#1.png}

\begin{document}


\title{Rule based End-to-End Learning Framework for Urban Growth Prediction}


\author[1]{Saptarshi Pal}
\author[1]{Soumya K Ghosh}

\affil[1]{(Department of Computer Science and Engineering, Indian Institute of Technology Kharagpur, India.)}


\maketitle

\begin{abstract}
Due to the rapid growth of urban areas in the past decades, it has become increasingly important to model and monitor urban growth in mega cities. Although several researchers have proposed models for simulating urban growth, they have been primarily dependent on various manually selected spatial and nonspatial explanatory features for building models. A practical difficulty with this approach is manual selection procedure, which tends to make model design process laborious and non-generic. Despite the fact that explanatory features provide us with the understanding of a complex process, there has been no standard set of features in urban growth prediction over which scholars have consensus. Hence, design and deploying of systems for urban growth prediction have remained challenging tasks. In order to reduce the dependency on human devised features, we have proposed a novel \textit{End-to-End} prediction framework to represent remotely sensed satellite data in terms of rules of a cellular automata model in order to improve the performance of urban growth prediction. Using our \textit{End-to-End} framework, we have achieved superior performance in \textit{Figure of Merit, Producer's accuracy, User's accuracy, and Overall accuracy} metrics respectively over existing learning based methods.
\end{abstract}


\section{Introduction}

Urban regions are fast growing entities in the modern world. Areas which have been outskirts of a city/town twenty years back are now habituated and developed into urban form. Research has shown that uncontrolled sprawl of urban areas has potentially adverse effects on the environment in terms of biodiversity, loss of habitat and change of landscape (\cite{basawaraja2011analysis, moghadam2013spatiotemporal}). United nations have predicted based on past data that urban population will rise from a current proportion of $54$ per cent to a proportion of $66$ percent by the year $2050$\footnote{http://www.un.org/en/development/desa/news/population/world-urbanization-prospects.html}. With the uncontrolled expansion of cities, it has become important to monitor and control changes in the urban landuse.

A fundamental difficulty in monitoring urban areas is due to its huge spread, which makes manual tracking of global landuse/built-up changes intractable. However, there are several remote sensing satellites periodically capturing images of the earth surface, which contains plenty of real time information regarding change of urban landscape. Hence, satellite imagery has been considered as an effective component for real time monitoring of urban growth.

Urban growth monitoring using satellite derived products has been addressed by several researchers (\cite{moghadam2013spatiotemporal, shafizadeh2015performance, lin2011predictive}). Models using cellular automaton, agent based methods, neural networks, logistic regression, fractals have been developed for robust and accurate prediction of urban growth (\cite{liu2008modelling, bhatta2012urban, musa2016review, aburas2016simulation}). Cellular automata is a widely preferred model for urban growth prediction due to its simplicity. Several CA models and corresponding case studies have been done on various major cities for detecting urban change, for instance, Vancouver Britain, Wuhan China etc (\cite{aburas2016simulation,rienow2015supporting, ke2016partitioned, van2009modeling, feng2016modeling}). Recently, artificial neural networks have also been quite popular in the land use change prediction (\cite{shafizadeh2015performance, razavi2014predicting}) and studies using the combination of neural networks and cellular automata models have been proposed (\cite{omrani2017integrating}). On the other hand, support vector machines (SVM), Decision trees and Random Forest based models have also been proposed (\cite{rienow2015supporting, ke2016partitioned, ahmadlou2016modeling, kamusoko2015simulating, li2004data}). Besides, logistic regression has been quite suitable for urban growth prediction in order to assess the driving forces behind urban growth (\cite{hu2007modeling}). Overall, among the learning techniques used for prediction, there have been broadly four categories which have dominated the urban growth community, namely logistic regression, artificial neural networks, support vector machines and inductive learning approaches (\cite{musa2016review}). 

Each of the models developed using learning methodologies mentioned above has utilized spatial and non-spatial explanatory features for predicting urban growth (\cite{musa2016review}). Spatial features such as distance from urban areas, water bodies, swamps etc. are derived from satellite imagery manually using supervised land use classification methods (\cite{shafizadeh2015performance, thapa2011urban}). Other kinds of spatial factors such as elevation, distance from roads and railways are derived from other open spatial data repositories like Open Street Maps\footnote{https://www.openstreetmap.org/} and US Shuttle Radar Topography Mission\footnote{https://lta.cr.usgs.gov/SRTM1Arc}. Three issues of this method are discussed as follows.

\begin{itemize}

\item It has been evident from practical experience (\cite{bengio2013representation}) that feature engineering significantly affects results of machine learning algorithms. At present, most of the urban growth prediction models use distance based features, for instance, distance from roads, city centers, railways, water bodies etc. and the distance metric is generally euclidean distance. Although the distance from certain regions does determine built-up, that may not be always euclidean distance. Moreover, there are other distance metrics, for instance, Minkowski's distance, Mahalanobis distance, etc which have not been sufficiently used in urban growth prediction modeling. It may be possible to achieve better performance using other distance metrics. The issue, in this case, is a selection of suitable distance based criterion which requires manual effort.

\item The spectral properties can also be a crucial feature as different spectral properties indicate different landuse and different landuse may have different growth patterns. Hence another limitation of distance based features is that they are unable to capture the spectral properties of landuse.

\item Due to the disparity in the geography of our earth surface and noise in satellite imagery, it is often excessively labor-intensive to derive all the different kinds of landuse feature maps from satellite imagery.

\end{itemize}


We intend to address these issues by designing an intelligent framework which has minimal human bias and build models in an \textit{End-to-End} fashion. The idea of \textit{End-to-End} learning is to bypass intermediate feature engineering steps in model building and design a model that can directly learn robust representations from input data (\cite{bengio2013representation, graves2014towards}). \textit{End-to-End} learning is a paradigm of machine learning that has recently gained importance due to the advent of \textit{Representation} learning and high-performance computing infrastructures. It has delivered exceptional performance in domains like image classification, speech processing, image segmentation etc (\cite{krizhevsky2012imagenet, badrinarayanan2015segnet, graves2013speech}). Majority of these techniques demonstrate the existence of complex relationships between various real world objects/phenomenon and conclusions that we draw from it. These complex relations are stored in our memory in terms of intricate knowledge structures which are hard to decode but makes human beings quite efficient in motor skills. 

Inspired by the above works, we put forward our hypothesis: \textit{Automatically extracted information from remotely sensed data can be an important feature in modeling urban growth.} The primary objective of this paper is to utilize the concept of \textit{Representation learning} in developing a novel \textit{End-to-End} framework for prediction of urban growth. \textit{The proposed framework utilizes unsupervised representation learning and supervised classification methods to learn rules of an urban growth cellular automata model from raw satellite data to form various knowledge structures.} Our proposed framework gives superior prediction performance than existing learning based methodologies in terms of various metrics which evaluates urban growth simulation performance.


The key contributions of this paper are as follows.

\begin{itemize}
\item Eliminating feature engineering module from prediction framework.

\item Generating representations from spectral information present in remote sensing satellite imagery in improving performance of urban growth prediction.
\end{itemize}

The rest of the sections are organized as follows. In section $2$, we define the cellular automata model for urban growth prediction. Section $3$ describes the framework that has been used to discover and store rules. Section $4$ presents results and discussions of the experiments conducted in the region of Mumbai, India. Finally, section $5$ provides conclusion and future research directions.

\section{Urban Growth modeled using Cellular Automata (CA)} \label{section:ugca}

Urban growth, being a spatiotemporal dynamic process, can be modeled using the theory of cellular automata. A typical CA model is composed of an infinite array of cells having a finite number of states, which transform at discrete time steps using certain transition rules. The transition rules of a CA model signify the relationship between a cell and its neighborhood. The neighborhood criterion can be Von Neumann ($4$ neighbors) or Moore Neighborhood ($8$ neighbors). Transition rules are iteratively applied to the cells to simulate dynamic processes over multiple time steps. 

We define a cell state at point $p$ and time $t$ as $S_p^t = <l_p^t, \tau_p^t>$, where $l_p^t$ is a binary label variable representing \textit{Built-up} and \textit{Non Built-up} and $\tau_p^t$ is a transition indicator. The range of label variable $l_p^t$ is $\{-1,1\}$, where $-1$ represents \textit{Non Built-up} and $1$ represents \textit{Built-up}. Let us represent the set of \textit{Built-up} and \textit{Non Built-up} pixels originally as $S_{B}$ and $S_{NB}$. The transition indicator indicates whether the cell under consideration has undergone a transformation in the current time step. The range of the transition indicator is $\{0,1,2,3\}$, where $0,1,2,3$ represents transition from  \textit{Non Built-up} to \textit{Non Built-up} ($C_{NB}^{NB}$), \textit{Built-up} to \textit{Built-up} ($C_{B}^{B}$), \textit{Non Built-up} to \textit{Built-up} ($C_{NB}^{B}$) and \textit{Built-up} to \textit{Non Built-up} ($C_{B}^{NB}$). A pictorial representation of the transition classes are displayed in Fig. \ref{fig:ca_decision_tree} (b).

We define the update rule for the transition indicator as 

\begin{eqnarray}
\tau_p^{t+1} = f_T(l_p^t, N(l_p^t))
\label{eqn:ca_update}
\end{eqnarray}

, where $N$ is the neighborhood criterion and $f_T$ is a update function to be modeled. The transition function for update of the state $S_p^{t}$ to $S_p^{t+1}$ given as is,

\begin{eqnarray}
S_p^{t+1} = 
\begin{cases}
(S_B,\tau_p^{t+1}) & \tau_p^{t+1}\in \{C_{NB}^B,C_B^B\} \\
(S_{NB},\tau_p^{t+1}) & \tau_p^{t+1}\in \{C_{NB}^{NB},C_{B}^{NB}\}
\end{cases} 
\label{eqn:ca_transition}
\end{eqnarray}

The theory of cellular automata is convenient in this scenario because it considers the space as a discrete array of cells. Since in our case, the data is of raster type, it is already in form of discrete matrix of pixels. So discretization of space is not required to be done explicitly. If the data was in vector form, then we would have to define the discretization  of the space separately.

\subsection{Extension of the CA model}

The urban growth CA model can be extended to include other variables which are driving factors in urban growth (\cite{yang2008cellular, feng2016modeling}). In order to insert the raster information in the CA model, we have modified our CA model in the following way.

\begin{itemize}
\item The state variable $S_p^t$ is extended to include a raster data variable $R_p$ from the images to form $<l_p^t, \tau_p^t, R_p>$. The raster image can consist of multiple bands in which case, $R_p$ represents a vector of values of all bands at point $p$.

\item The update equation (\ref{eqn:ca_update}) gets transformed to include the raster variable $R_p$ in the following way.

\begin{eqnarray}
\tau_p^{t+1} = f_T(l_p^t, N(l_p^t), \phi_{raster}^{len}(R_p, N(R_p)))
\label{eqn:ca_update_raster}
\end{eqnarray}
\end{itemize} 

The above modification allows us to reflect upon the relationship between the \textit{built-up} variables and the raster variables through the function $f_T$ and $\phi_{raster}$. The function $\phi_{raster}^{len}$ is an encoding function which is responsible for generating a desired length $len$ encoding of the information present in $<R_p, N(R_p)>$. Thus, we have reduced the problem of modeling urban growth from raster data to modeling the function $f_T$ and $\phi_{raster}^{len}$.

\input{endtoend.tex}

\input{exp.tex}

\section{Conclusion and Future Work}

We have introduced the concept of  \textit{End-to-End} learning in urban growth prediction by developing a framework for learning rules of a cellular automata model directly by representing spectral information from remote sensing data. We have empirically verified our framework on predicting urban growth for the region of Mumbai over a time frame of $20$ years. Our \textit{End-to-End} framework has outperformed existing learning based methodologies with a simpler implementation than the existing frameworks. 

Future work can be based on challenges which we have encountered in this work. Since spatial resolution of the satellite is fixed, therefore as temporal resolution reduces, the number of cells in which built-up happens reduces. Due to this, the number of points in the transition classes reduces and persistence classes increases. Therefore, the dataset gets heavily imbalanced and the ability of the classifier to learn patterns of transition classes reduces. The problem of imbalanced datasets can be alleviated if both the temporal and spatial resolution of the satellite images are increased. This is possible if data from high-resolution satellites like IKONOS/ QuickBird are used. However, the data load of high-resolution satellites is high, hence more infrastructure and sophisticated algorithms would be necessary.

Despite superior performance of the \textit{End-to-End} approach, one of the drawbacks of \textit{End-to-End} learning is essentially difficulty in understanding the rules. Hence, uncovering automatically generated rules in an \textit{End-to-End} learning framework can be considered as a challenge which needs to be resolved in the subsequent studies. Furthermore, the \textit{End-to-End} framework can be extended to incorporate other data resources which are vector form, which may further improve the performance of the framework.

\input{main.bbl}

\end{document}

%% file: endtoend.tex
\section{End-to-End Framework for learning Cellular Automata rules}

In this section, we discuss the details of our architecture, illustrated in Fig. \ref{fig:ca_decision_tree} (a), for generating a transition function $f_T$ for the cellular automata model described in eqn (\ref{eqn:ca_update_raster}). The architecture comprises of the following components.

\begin{itemize}
\item Two data repositories consisting of built-up rasters and remotely sensed rasters which will be required during model building. 

\item Data representation module is for representing the data obtained from the data repositories in such a way so as to facilitate learning and knowledge representation.

\item The learning algorithm stage is expected to learn patterns from the data which is received from the previous stage.

\item The knowledge representation module stores the knowledge learned by the learning algorithm for further querying.

\item Finally, the prediction stage predicts future built-up conditions. 
\end{itemize}

\subsection{Data Representation} \label{section:data_rep}

The data representation component consists of procedures to build a \textit{data matrix} and a \textit{label matrix} from the raster and built-up maps. The \textit{data matrix} is formed from the raster and \textit{built-up} maps in form of a matrix of size $n\times (1+|N(l_p^t)|+|\phi_{raster}^{len}(R_p, N(R_p))|)$, where $n$ is the number of data points in the raster, $N()$ is a neighborhood function and $|.|$ is the cardinality of a set. The neighborhood set $N()$ plays a significant role in the framework, as it represents the effect of neighboring pixels. Increase in the neighborhood set size increases the model complexity by increasing the dimensions of the \textit{data matrix} (\cite{sante2010cellular}).

\begin{figure}
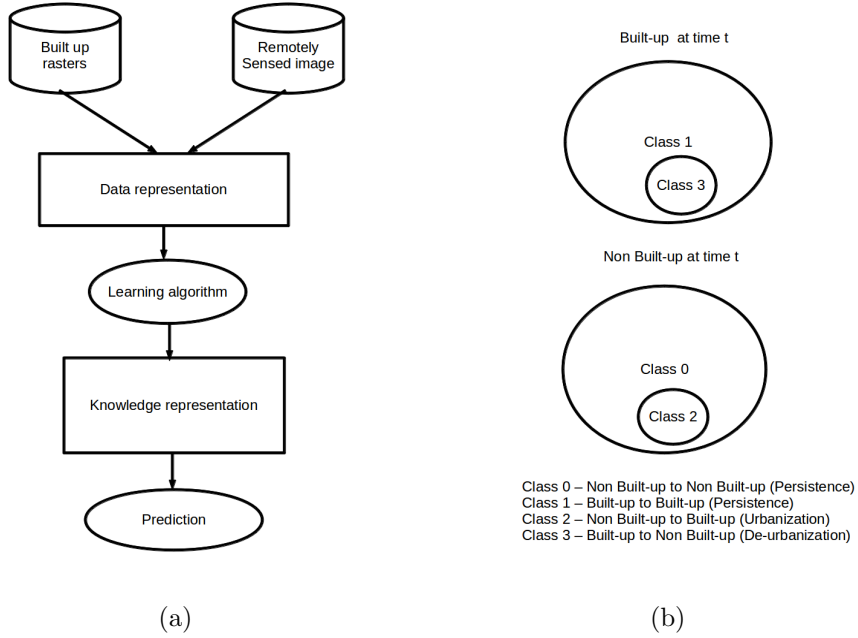

\centering
\subfloat[]{\includegraphics[scale=0.3]{\locimage{ca_decision_tree}}}
\subfloat[]{\includegraphics[scale=0.3]{\locimage{transition_classes}}}
\caption{(a) \textit{End-to-End} framework for prediction of urban growth (b) Venn diagram of transition classes relevant to Urban growth prediction}
\label{fig:ca_decision_tree}
\end{figure}

\begin{figure}
\centering
\includegraphics[scale=0.6]{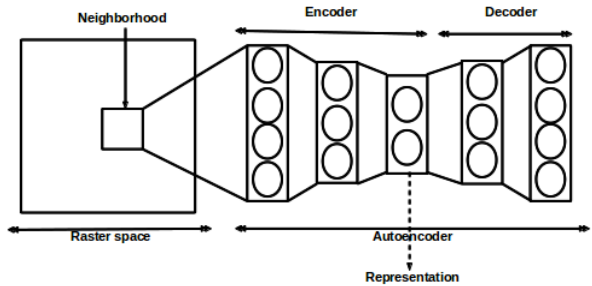}
\caption{Learning Representation from Neighborhood using an Autoencoder.}
\label{fig:rasterrep}
\end{figure}

A crucial step at this stage is the design of the encoder function $\phi_{raster}^{len}$. We can build an encoder function by training an autoencoder\cite{vincent2010stacked} with the vectors of form $<R_p, N(R_p)>$ to form a $len$ length representation. An autoencoder is a neural network consisting of an encoder which encodes the data vector of a certain size and a decoder which reconstructs the data vector back from the encoding. The network is trained by minimizing the error between the decoded output and the original output. Finally the encoder from the autoencoder is used to encode the input which is known as a representation. Since the method is unsupervised, therefore there is no requirement to label the data for encoder design. The encoder design for our framework is depicted in Fig. \ref{fig:rasterrep}. If a vector is represented by $X = <R_p, N(R_p)>$, then $\phi_{raster}^{len}$ can be expressed as

\begin{eqnarray}
\phi_{raster}^{len}(X) = \sigma(W_k^e...\sigma(W_1^eX))
\label{eqn:encoder_raster}
\end{eqnarray}

, where $W_k^e$ is a weight matrix and $\sigma$ is an activation function (eg. sigmoid, relu). The eqn. \ref{eqn:encoder_raster} represents a $k$ layer feed forward neural network which encodes the information in a raster position and its neighborhood. Similar to the encoder function, we can express the decoder function (say $\phi'$) as

\begin{eqnarray}
\phi'(X) = \sigma(W_k^d..\sigma(W_1^d \phi_{raster}^{len}(X)))
\label{eqn:decoder_raster}
\end{eqnarray} 

, where $W_k^d$ is a weight matrix and $\sigma$ is an activation function. The only constraint in eqn. \ref{eqn:decoder_raster} is that the output vector $\phi'(X)$ must have the equal dimension to that of $X$. Then the optimization of the parameters can be done using the following minimization equation.

\begin{eqnarray}
E = \frac{1}{n} \sum{(X-\phi'(X))^2}
\label{eqn:optimization}
\end{eqnarray}

Subsequent to the optimization, the representations of length $len$ can be generated from the vectors $<R_p, N(R_p)>$ using the function $\phi_{raster}^{len}$ as in eqn. \ref{eqn:encoder_raster}.

The \textit{data matrix} can be obtained by concatenating the \textit{built-up} vectors ($<l_p^t,N(l_p^t)>$) and \textit{raster representation} ($<\phi_{raster}^{len}(R_p, N(R_p))>$) columnwise. The \textit{raster representation} variables are non-categorical with values ranging from $[-1,1]$, whereas the \textit{built-up} variables are categorical with value either $-1$ (Non Built-up) or $1$ (Built-up). The rows of the \textit{data matrix} can be directly supplied to the function given in eqn. (\ref{eqn:ca_update_raster}). 

The \textit{label matrix} is an $n\times 1$ matrix which consists of labels of $4$ transition classes namely, \textit{Non Built-up} to \textit{Non Built-up}, \textit{Non Built-up} to \textit{Built-up}, \textit{Built-up} to \textit{Non Built-up} and \textit{Built-up} to \textit{Built-up}. The transition classes and their class labels are illustrated as a Venn diagram in Fig \ref{fig:ca_decision_tree} (b). Classes $2$ and $3$ are within classes $0$ and $1$ because in the dataset usually number of transition cells are lesser in number than number of persistent cells. The rows of the \textit{data matrix} form the input vector to the function $f_T$, whereas rows of the label matrix form output values of the function $f_T$. The procedure for building the \textit{data and label matrix} is given in Algorithm \ref{algo:dataprep}.

Contrary to other works (\cite{moghadam2013spatiotemporal, shafizadeh2015performance, yang2008cellular}), where two classes \{\textit{Built-up, Non Built-up}\} have been used for training classifiers, we have used four classes in our \textit{label matrix}. This resolves two following issues.

\begin{itemize}

\item With the two class approach, the framework turns into a built-up classification problem because the raster images consist of information regarding built-up. In our case, we have four classes out of which two are transition and two are persistence classes. This tweak divides the \textit{data matrix} into four parts and patterns of transition and persistence classes get separated. The classifier still does classification, but the pattern classification problem turns into, ``Which patterns lead to transition and which ones lead to persistence?".

\item Comparing the \textit{Built-up} maps of two close time steps, we have observed that transition from \textit{Non Built-up} to \textit{Built-up} and \textit{Built-up} to \textit{Non Built-up} occurs in a small fraction of cells when compared to cells which are persistent. In the $2$-class approach of modeling, certain learning algorithms tend to ignore patterns that are less fraction of the dataset considering them as noise. For instance, if there are $10\%$ transition pixels and $90\%$ persistent pixels, then a default choice for high predictive accuracy is to give an accuracy of $90\%$ (\cite{chawla2002smote}). The imbalance in the dataset is unavoidable in this case as it is a natural property of the problem at hand, which gets covered up when not considered separately. Imbalanced datasets are common in many practical problems and have been addressed multiple times (\cite{chawla2005data}). In our case, we have used three separate metrics, namely \textit{Figure of Merit (FoM), Producer's Accuracy (PA)} and \textit{User's Accuracy (UA)} other than classification accuracy to track whether this problem is tackled by the knowledge representation or not (\cite{pontius2008comparing}). These three metrics are designed to check whether a landuse change model can predict the transitions properly. Therefore, it is crucial to have high FoM, PA and UA along with OA to guarantee a good simulation technique. 

\end{itemize}

\subsection{Knowledge Representation}

\textit{Knowledge as defined by Fischler and Firschein 1987, refers to stored information or models used by person or machine to interpret, predict, and appropriately respond to the outside world}. By the very nature of the definition, knowledge is task specific (\cite{haykin2009neural}). There are various ways in which knowledge can be represented such as rules (Decision Trees), decision surfaces (SVM), computation nodes (ANN), probability distributions (Probabilistic Graphical models) and nearest neighbors (KNN). Table \ref{table:know_tech} provides a description of the various kinds of supervised classification methods, which can be considered in the \textit{End-to-End} framework for knowledge representation (Only short mathematical descriptions are provided in the table). The reason behind referring these as knowledge representation techniques is as follows.

A classifier is primarily comprised of a mathematical form and a parameter estimation (training) technique. The mathematical form consists of a set of parameters which represents a family of functions. The parameter estimation technique is an algorithm for finding a suitable set of parameters corresponding to the data, that optimizes a particular objective. When a classifier is trained using data, it builds a knowledge structure based on its corresponding mathematical representation. The developed knowledge represented by a classifier is the set of parameters, which have been obtained after parameter estimation. The performance of the knowledge representation technique on a particular problem depends on how precise the set of parameters is estimated from the data.

To the best of our knowledge, the selection of a classifier requires a profound understanding of the data, which may not be always available for complex multivariate problems. Since any of the classifiers in Table \ref{table:know_tech} can represent the transition function $f_T$ (given in eqn. (\ref{eqn:ca_update_raster})), therefore one needs to select a classifier based on certain criterion.

In our case, the selection of the classifier as a knowledge representation technique is based on eight metrics, namely \textit{Cross validation, Training time, Prediction time, Figure of Merit (FoM), Producer's accuracy (PA), User's accuracy (UA)} and \textit{Overall accuracy (OA)}. Each of these metrics is a different window that provides a certain view of the framework --- \textit{Cross validation} measures overfitting; \textit{FoM, PA, UA} and \textit{OA} measure framework's performance in learning patterns and simulation; \textit{Training} and \textit{Prediction time} measure swiftness of the knowledge representation technique. Based on all the views, we have selected our suitable knowledge representation technique as a Decision tree or Ensemble of Decision trees (Random Forests) for generation of cellular automata rules. Hence we call the framework a \textit{Rule based End-to-End} framework.

\begin{table}
\centering
\resizebox{\textwidth}{!}
{
\begin{tabular}{|c|c|c|c|}
\hline
{\bf Classifier} & {\bf Knowledge representation} & \specialcell{\bf Mathematical representation} & {\bf Training procedure}\\
\hline
Logistic regression & \specialcell{Stores Knowledge\\ in terms of\\ weights ($W$ and $b$)\\ of a hyperplane} & $Y = \sigma(W*X+b)$ & Stochastic gradient descent \\
\hline
Gaussian Naive Bayes & \specialcell{Stores knowledge\\ in terms of\\ Probability distribution} & $Y = \operatorname*{arg\,max}_{k\in \{1,2,..K\}} p(C_k) \prod_{i=1}^{N} P(x_i|C_k)$ & Maximum likelihood estimation\\
\hline
Support vector machines & \specialcell{Stores knowledge in\\ terms of weights\\ ($W$ and $b$) and kernel $\phi$} & $W*\phi(X)-b=0$ & Sequential minimal optimization\\
\hline
Multi Layer Perceptrons & \specialcell{Stores knowledge\\ in terms of\\ computation nodes} & $\sigma(W_1*\sigma(..\sigma(W_n * X)))$ & Backpropagation\\
\hline
K-Nearest Neighbors & \specialcell{Stores feature vectors \\and class labels \\as knowledge} & \specialcell{feature vectors and\\ class labels} & \specialcell{instance based learning\\ or non-generalizing learning}\\
\hline
Decision Tree & \specialcell{Stores knowledge\\ in terms\\ of rules} & $(X_i = a_i) \wedge ... \wedge (\theta_j = a_j)$ & CART, ID3, C4.5\\
\hline
\specialcell{Ensemble methods\\ (Random forest, AdaBoost)} & \specialcell{Stores knowledge\\ in terms of the\\ unit classifier \\in the ensemble} & Depends on the unit classifier & Bagging/Boosting\\
\hline
\end{tabular}
}
\caption{Features of popular Classifiers which are also knowledge representation technique.}
\label{table:know_tech}
\end{table}

\subsection{Methodology}

As discussed in Section \ref{section:ugca}, the Urban Growth Cellular Automata model consists of three components namely, cell state $S_t^p$, transition function $f_T$ and update rule (eq. (\ref{eqn:ca_transition})). The cell state and update rule needs to be predefined and is not flexible, while the transition function is capable of taking any form during the training. Hence the framework is generic and does not depend on manually designed features. The procedures for dataset preparation, training and prediction are given in the Algorithms \ref{algo:dataprep}, \ref{algo:train} and \ref{algo:predict} respectively. 

Algorithm \ref{algo:dataprep} consists of the following steps.

\begin{itemize}

\item Gathering raster and built-up values of a point and its neighborhood in a matrix $X_R$ with row vector ($x_r \gets \{R(p), N(R(p)\}$) and in a matrix $X_B$ with row vector ($x_b \gets \{B_p^t, N(B_p^t)\}$) respectively. The matrix $X_R$ is used to build an autoencoder using a custom function $build\_autoencoder()$. This is followed by generation of representations ($\phi_{raster}^{len}(x_r)$) and collection of all points in $X_B$ and representations $\phi_{raster}^{len}(x_r)$ to form a data matrix. It should be noted that the neighborhood criterion needs to be decided before the Data generation process. In this paper, we have used a standard Moore neighborhood criterion.

\item Preparation of label matrix which is composed of transition and persistent classes. To prepare the matrix, we have declared \textit{isUrban($B_p^t$)} function, which returns \textit{true} if the point $p$ at time $t$ is urban and \textit{false} otherwise. It is required to determine whether at point $p$ and time interval $\{t, t+1\}$ transition has occurred or not.

\end{itemize}

Algorithm \ref{algo:train} consists of the following steps.

\begin{itemize}

\item Retrieve the \textit{Data matrix} and \textit{Label matrix} from Algorithm \ref{algo:train}.

\item Select a set of classifiers $S_C$ as in Table \ref{table:know_tech} and train these classifiers using their corresponding training algorithm.

\item The resultant is a set of trained classifiers, each of which can be considered as a transition function $f_T$.

\item Then, depending on the Figure of Merit, Producer's accuracy, User's accuracy, Overall accuracy, cross-validation, training time and prediction time, one can choose which one to use. In certain cases, it may be convenient to use cross-validation, training time and prediction time to remove certain classifiers from the list initially.

\end{itemize}

Algorithm \ref{algo:predict} consists of the following steps.

\begin{itemize}

\item Gathering raster and built-up values of a point and its neighborhood in row vectors $x_r$ and $x_b$ respectively.

\item Using $\phi_{raster}^{len}$ from Algorithm \ref{algo:dataprep} to generate representations. This is followed by concatenation of the vectors $x_b$ and representations $\phi_{raster}^{len}(x_r)$ to form $x$.

\item Use a trained classifier $f_T$ to predict on $x$.

\item If $f_T$ predicts a transition from \textit{Non Built-up} to \textit{Built-up} or a persistence from \textit{Built-up} to \textit{Built-up}, then the final class at $t+1$ is \textit{Built-up} and \textit{Non Built-up} otherwise.

\end{itemize}

\begin{figure}[t]
\centering
\includegraphics[scale=0.25]{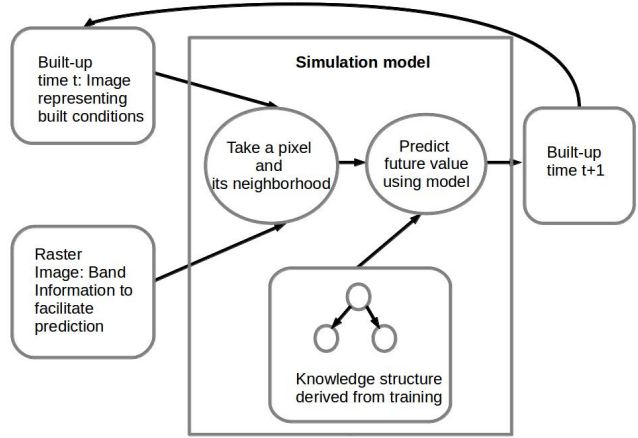}
\caption{Recurrence of the Rule based Framework due to the cellular automata}
\label{fig:recurrent_framework}
\end{figure}

It may be noted that at least one built-up image is necessary as an initial point to start prediction procedure. This is due to the fact that we have modeled urban growth as a cellular automaton for which the framework is recurrent in nature (see Fig. \ref{fig:recurrent_framework}). Figure \ref{fig:recurrent_framework} represents the flowchart of the simulation procedure. It shows that the framework can simulate urban growth up to as many years as possible in the future each time utilizing the last predicted image.

It has been proved in (\cite{omohundro1984modelling}) that \textit{an arbitrary $9$ neighbor, two-dimensional cellular automata can be simulated in terms of a set of ten partial differential equations}. The PDEs indicate the relation between the dynamic variable and space-time, which can be a theoretical form of the urban growth cellular automata. Without the cellular automata, the procedure would be merely a built-up/landuse classification rather than prediction.

\begin{algorithm}
\begin{algorithmic}
\STATE {\bf Input:}
\STATE $B^t \gets \textit{Built-up raster at time t}$
\STATE $B^{t+1} \gets \textit{Built-up raster at time t+1}$
\STATE $R \gets \textit{Raster image}$
\STATE $isUrban()\gets$ function which returns True if point is Urban.
\STATE $build\_autoencoder() \gets$ function which builds a function $\phi_{raster}^{len}$ after training on a dataset.
\STATE {\bf Output:}
\STATE $X, Y, \phi_{raster}^{len} \gets$  \textit{Data matrix}, \textit{Label matrix} and \textit{Encoder function}.
\STATE {\bf Procedure:}
\STATE $X, Y, X_R, X_B\gets \phi$ 
\FORALL {pixel $p$}
	\STATE $x_b \gets \{B_p^t, N(B_p^t)\}$  
	\STATE $x_r \gets \{R(p), N(R(p))\}$
	\STATE $X_B \gets X_B\cup x_b$
	\STATE $X_R \gets X_R\cup x_r$
	\IF{!isUrban($B_p^{t+1}$) and !isUrban($B_p^{t}$)}
		\STATE $Y \gets Y\cup C_{NB}^{NB}$
	\ELSIF{isUrban($B_p^{t+1}$) and isUrban($B_p^{t}$)}
		\STATE $Y \gets Y\cup C_{B}^{B}$
	\ELSIF{isUrban($B_p^{t+1}$) and !isUrban($B_p^{t}$)}
		\STATE $Y \gets Y\cup C_{B}^{NB}$
	\ELSE
		\STATE $Y \gets Y\cup C_{NB}^{B}$
	\ENDIF	
\ENDFOR
\STATE $\phi_{raster}^{len} \gets build\_autoencoder(X_R, len)$
\STATE $X \gets X \cup \{x_b,\phi_{raster}^{len}(x_r)\}; \forall x_r,x_b\in X_R,X_B$
\STATE return $X,Y,\phi_{raster}^{len}$
\end{algorithmic}
\caption{Data Representation procedure}
\label{algo:dataprep}
\end{algorithm}

\begin{algorithm}
\begin{algorithmic}
\STATE {\bf Input:}
\STATE $X \gets$ \textit{Data Matrix}
\STATE $Y \gets$ \textit{Label Matrix}
\STATE $S_C \gets$ Set of classifiers as in Table \ref{table:know_tech}
\STATE $C.train() \gets$ Trains $C$ with corresponding training algorithm
\STATE {\bf Output:}
\STATE $S_C^T \gets$ Set of trained classifiers on $X$ and $Y$
\STATE {\bf Procedure:}
\STATE $S_C^T \gets \phi$
\FORALL {$C\in S_C$}
	\STATE $C.train(X,Y)$ 
	\STATE $S_C^T = S_C^T\cup C$
\ENDFOR
\STATE return $S_C^T$
\end{algorithmic}
\caption{Parameter estimation procedure}
\label{algo:train}
\end{algorithm}

\begin{algorithm}
\begin{algorithmic}
\STATE {\bf Input:}
\STATE $B^t \gets \textit{Built-up raster at time t}$
\STATE $R \gets \textit{Raster image}$ 
\STATE $f_T \gets C \in S_C^T$ (from Procedure \ref{algo:train}) 
\STATE $\phi_{raster}^{len} \gets $ Encoder function from the output of Algorithm \ref{algo:dataprep}.
\STATE {\bf Output:}
\STATE $B_{t+1} \gets$ Future Built-up map.
\STATE {\bf Procedure:}
\FORALL {pixel $p$}
	\STATE $x_b \gets \{B_p^t, N(B_p^t)\}$
	\STATE $x_r \gets \{R(p), N(R(p)\}$   
	\STATE $x \gets \{x_b, \phi_{raster}^{len} (x_r)\}$
	\IF{$f_T(x) \in \{C_{NB}^B, C_B^B\}$}
		\STATE $B_p^{t+1} \gets$ Built-up
	\ELSE
		\STATE $B_p^{t+1} \gets$ Non Built-up
	\ENDIF
\ENDFOR
\STATE return $B^{t+1}$
\end{algorithmic}
\caption{Prediction procedure}
\label{algo:predict}
\end{algorithm}

\subsection{Features of the End-to-End Framework}

The proposed \textit{End-to-End} architecture has certain key features which differentiates it from earlier proposed urban growth prediction frameworks (\cite{shafizadeh2015performance, lin2011predictive, liu2008modelling, van2009modeling}). 

\begin{itemize}

\item The key improvement is the removal of a manual feature extraction module, which is inherent in the existing architectures and replacing it with a data representation module. The data representation module generates representations from the raw satellite images in an unsupervised manner in a form as described in Section \ref{section:data_rep} and does not require any separate database consisting of manually selected explanatory features.

\item The knowledge representation layer stores knowledge that directly relates built up and raster representations with the transition classes. This is a significant distinction from the earlier models where learning models were used to establish relationships between manually selected explanatory features and \textit{Built-up}. The removal of manual selection processes reduces the bias on models and knowledge structures, thereby creating an opportunity to develop new theories and explain the results of a complex process as urban growth.

\item Furtheremore, since representation learning is unsupervised, therefore it can be done without taking into account the final objective (in this case urban growth). Thus we can say that the representations are generic in nature and can be used for other applications as well.

\end{itemize}

Finally, these qualities come with a drawback, as implementation becomes easier but knowledge representation turns incomprehensible. Hence theoretically, it becomes a challenging task to extract meaning from these rules. This is due to the fact that the variables in the \textit{Data matrix} are frequency/band values which can have multiple semantics. Therefore, the rules generated can have multiple interpretations among which the most appropriate interpretation needs to be found out.

%% file: exp.tex
\section{Experiments and Results}

We have conducted our experiments on the region of Mumbai, which is the capital city of Maharashtra, India (latitude: $19.0760\degree$ N, longitude: $72.8777\degree$ E). The city lies on the west coast of India and has a natural harbor. Mumbai is one of the mega cities of India also often colloquially known as the City of Dreams. According to Census India, the population of Mumbai has been steadily rising from approximately $9$ million in $1991$ to more than $12$ million in $2011$. The region under consideration is shown in Fig. \ref{fig:builtup_rasters}a.

The experiments are conducted in a Virtual Machine (VM) of an OpenStack based cloud infrastructure\footnote{http://www.sit.iitkgp.ernet.in/Meghamala} with the \textit{8 VCPUs}, \textit{16 GB RAM} and operating system \textit{Ubuntu 14.04}.

\subsection{Data Collection and Preprocessing}

We have collected remotely sensed natural color ($3$ bands) Landsat images from the United States Geological Survey (USGS) website\footnote{http://glovis.usgs.gov/} for the years $1991$, $2001$ and $2011$ for the Mumbai region. The Mumbai region has been extracted from the raster and the final image consisted of $972280$ data pixels. The images have been segmented to generate the \textit{built-up} raster images that are binary images with white pixels representing \textit{built-up} and black pixels representing \textit{non built-up}. The segmentation has been carried out using maximum likelihood classification implemented in semi-automatic classification plugin\footnote{https://plugins.qgis.org/plugins/SemiAutomaticClassificationPlugin/} of QGIS\footnote{http://www.qgis.org/en/site/}. The semi-automatic classification method is essentially a semi-manual labeling method where initial labels are to be provided by a human and then segmented maps are generated using the raster values of the Landsat image. Since it is manual and inaccurate it needs to be verified to some source or reference maps which in our case have been Google Earth and Mumbai maps from a previously published work on Mumbai \cite{shafizadeh2015performance}. The total number of pixels that transformed from \textit{non built-up} to \textit{built-up} during the $20$ years considered in our study are given in Table \ref{table:pixel_transformed}. The percentage of pixels that changed from non-urban to urban is approximately $10\%$, which is similar to other studies conducted on the region of Mumbai (\cite{shafizadeh2015performance, moghadam2013spatiotemporal}). The slight aberrations are due to the classification inaccuracy of the classifiers used for performing the landcover classification.

\begin{table}[h]
\centering
\begin{tabular}{|c|c|c|}
\hline
Time step & Pixels transformed & Pixels persistent\\
\hline
$1991-2001$ & $149746$ & $822534$\\
\hline
$2001-2011$ & $103360$ & $868920$\\
\hline
\end{tabular}
\caption{Number of pixels transformed vs number of pixels persistent}
\label{table:pixel_transformed}
\end{table}

\begin{figure*}
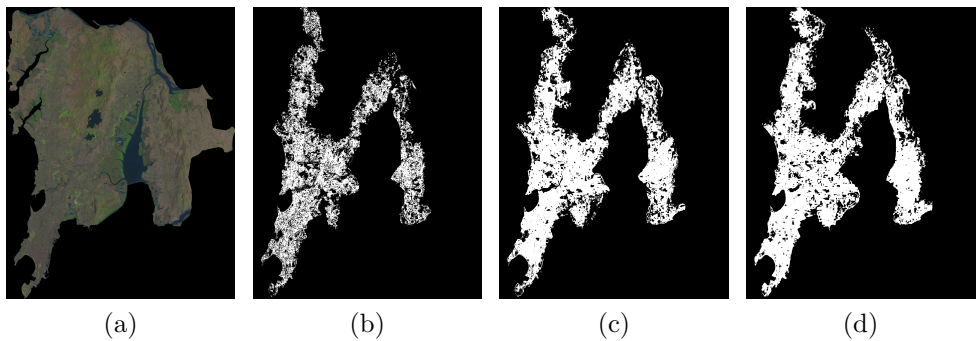

\centering
\subfloat[][]{\includegraphics[scale = 0.07]{\locimage{mumbai1991}}}\hspace{0.1cm}
\subfloat[][]{\includegraphics[scale = 0.07]{\locimage{cimg1991}}}\hspace{0.1cm}
\subfloat[][]{\includegraphics[scale = 0.07]{\locimage{cimg2001}}}\hspace{0.1cm}
\subfloat[][]{\includegraphics[scale = 0.07]{\locimage{cimg2011}}}
\caption{Rasters showing (a) Mumbai year 1991 and \textit{built-up} conditions of the year (b) 1991 (c) 2001 (d) 2011. (White represents built-up while black represents non built-up)}
\label{fig:builtup_rasters}
\end{figure*}

The rasters showing the urban growth conditions of the years $1991$, $2001$ and $2011$ are shown in Fig \ref{fig:builtup_rasters}. In our case, we have considered $3$ classes i.e. $C_{NB}^{NB}, C_{NB}^{B}$ and $C_B^B$. The $4^{th}$ class $C_{B}^{NB}$ have had $0.02$ per cent contribution in the training set, probably as there have been no recent massive de-urbanization in the area. Hence we have considered all such instances to also fall in the class $C_{B}^{B}$. The \textit{data matrix} and the \textit{label matrix} are generated from the data considering a Moore neighborhood of radius $1$. The length of the encoded representation is varied in multiples of $5$. The autoencoder is trained using the Adaptive Gradient Descent Algorithm \cite{duchi2011adaptive} with batch size of $1000$. Subsequent to the training, the encoder present in the autoencoder is used to generate representations for creating the \textit{data matrix}.

\subsection{Training and Validation}

The \textit{data matrix} and \textit{label matrix} generated in the previous step have been used to develop knowledge structures using various kinds of knowledge representation techniques as given in Table \ref{table:know_tech}. The classifiers have been trained in a multi-parameter setting in the following ways.

\begin{itemize}

\item \texttt{Logistic Regression}: For this classifier, we have executed the training for $10$ times each with a values of L2 regularization in the range $[0.01,100]$.

\item \texttt{Gaussian Naive Bayes}: No special prior probabilities have been set for this classifier.

\item \texttt{Support Vector Machine}: Different kinds of kernel functions like linear, polynomial and radial basis function have been used to test performance metrics. For polynomial kernels, the degree of polynomial have been varied from $1$ to $10$. 

\item \texttt{Multi Layer Perceptron}: Parameters like number of hidden layers, hidden layer sizes, batch size, number of iterations, learning rate, momentum have been varied in different ranges to predict performance. The hidden layer configurations which have been tested are $(10,)$, $(20,15)$, $(20,15,10)$, $(20,15,10,5)$, $(20,15,10,5,3)$. The number of iterations and batch size have been taken in multiples of $100$. Learning rate have been taken with constant, adaptive and inverse scaled and in the range $[0.001, 0.1]$. Momentum have been varied in the range $[0.5,0.9]$.

\item \texttt{Single and Ensemble of decision trees}: For this classifier, we have used Gini impurity as the measure to separate the datasets. The maximum height of the tree is set to values in the range $\{10,100,200,\infty\}$. The algorithm used to build the decision tree is an optimized version of CART (Classification and Regression Trees), which can handle both categorical and non-categorical data\footnote{http://scikit-learn.org/stable/modules/tree.html\#tree-algorithms}. For ensemble method of knowledge representation, we have considered $\{10,100,1000\}$ decision trees/estimators. The implementations of the learning methods that we have used are available at scikit learn\footnote{http://scikit-learn.org/} library in python.

\end{itemize}

The average of all the results from all the different kinds of parameter settings are taken as the final performance of a knowledge structure.

According to \cite{pontius2008comparing}, only comparing classification accuracy is not enough for validating a landuse change model. They argued that in order to validate a change model, there needs to be a validation of $4$ metrics, namely Figure of Merit (FoM), Producer's accuracy (PA), User's Accuracy (UA) and Overall Accuracy (OA). Since urban growth is a land use change model, therefore we have used these $4$ metrics for model validation. The former $3$ metrics assist in comparing the original and predicted maps of urban growth, while Overall Accuracy (OA) can be thought of as classification accuracy. According to (\cite{pontius2008comparing}), the $4$ validation measures depend on $5$ variables which are 

\begin{itemize}
\item $A$ = Area of error due to observed change predicted as persistence.

\item $B$ = Area correct due to observed change predicted as change.

\item $C$ = Area of error due to observed change predicted in the wrong gaining category.

\item $D$ = Area of error due to observed persistence predicted as change.

\item $E$ = Area correct due to observed persistence predicted as persistence.
\end{itemize}

Figure of Merit ($FoM$) provides us the amount of overlap between the observed and predicted change. Producer's accuracy ($PA$) gives the proportion of pixels that model accurately predicted as change, given that the reference maps indicate observed change. User's accuracy ($UA$) gives the proportion of pixels that the model predicts accurately as change, given that the model predicts change. The equations of the metrics are given as follows.

\begin{eqnarray}
FoM = \frac{B}{A+B+C+D}
\label{eqn:FoM}
\end{eqnarray}

\begin{eqnarray}
PA = \frac{B}{A+B+C}
\label{eqn:PA}
\end{eqnarray}

\begin{eqnarray}
UA = \frac{B}{B+C+D}
\label{eqn:UA}
\end{eqnarray}

\begin{eqnarray}
OA = \frac{B+E}{A+B+C+D+E}
\label{eqn:OA}
\end{eqnarray}

\begin{figure*}
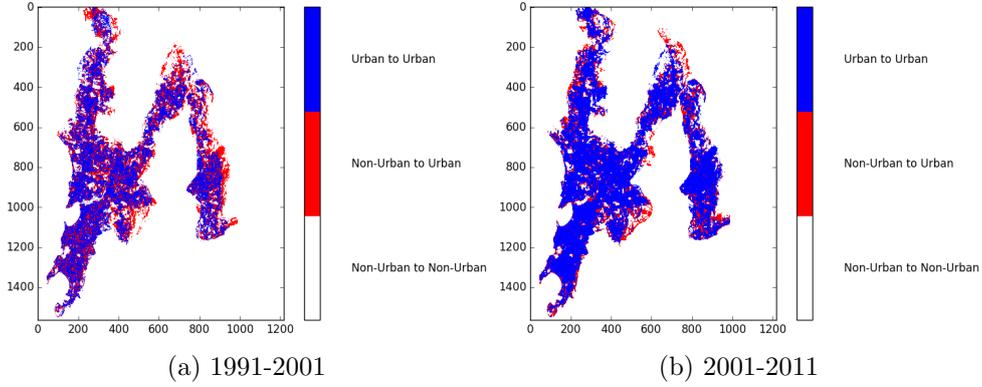

\centering
\subfloat[][1991-2001]{\includegraphics[scale = 0.34]{\locimage{mumbai-pred-1991-2001-DT}}}
\subfloat[][2001-2011]{\includegraphics[scale = 0.34]{\locimage{mumbai-pred-2001-2011-DT}}}
\caption{Predicted maps of Mumbai using Decision Tree (a) $1991-2001$ (b) $2001-2011$}
\label{fig:predmaps_DT}
\end{figure*}

\begin{figure*}
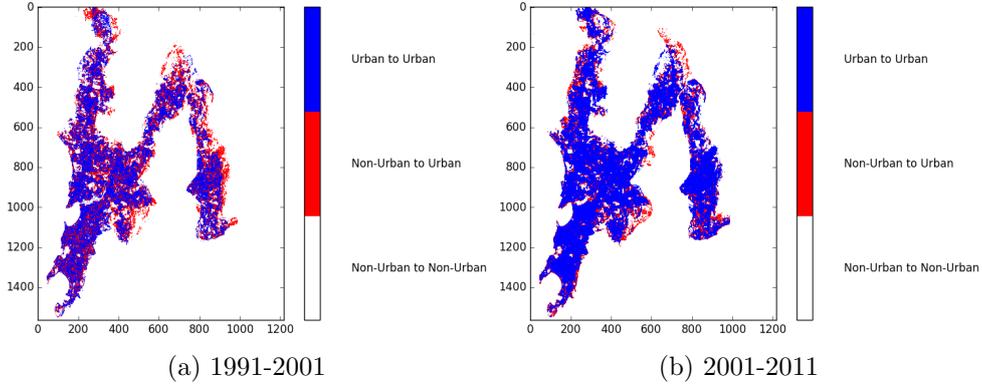

\centering
\subfloat[][1991-2001]{\includegraphics[scale = 0.34]{\locimage{mumbai-pred-1991-2001-RF}}}
\subfloat[][2001-2011]{\includegraphics[scale = 0.34]{\locimage{mumbai-pred-2001-2011-RF}}}
\caption{Predicted maps of Mumbai using Random Forest (a) $1991-2001$ (b) $2001-2011$}
\label{fig:predmaps_RF}
\end{figure*}

The training and validation of \textit{End-to-End} framework have been done using \textit{cross validation}. \textit{Cross validation} by dividing the dataset (generated from the dataset built from the years $1991-2001$) into $10$ parts and randomly selecting a part as validation set while others are used for training. The reason for selecting this is because the number of pixels which changed during this period is more than the other periods (Table \ref{table:pixel_transformed}). The validation results in terms of classification accuracy mean and variance is shown in Table \ref{table:crossval}. Performance comparison of existing methods with \textit{End-to-End} approach is presented in Fig \ref{fig:comparison_trset} (a) and (b). The first $4$ set of bars presents the results of existing methods applied to our dataset, whereas the last two set of bars provides the results of \textit{End-to-End} approach with single and ensemble of decision trees. The resultant built up maps representing transition classes predicted by both our approaches are displayed in Fig. \ref{fig:predmaps_DT} and \ref{fig:predmaps_RF}.

\begin{figure}
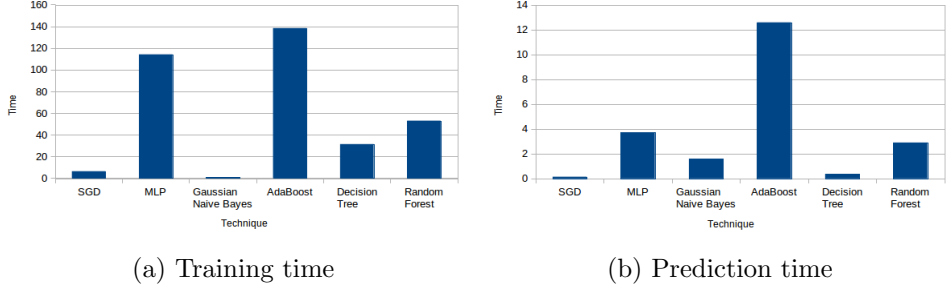

\centering
\subfloat[][Training time]{\includegraphics[scale=0.4]{\locimage{time_training}}}
\subfloat[][Prediction time]{\includegraphics[scale=0.4]{\locimage{time_test}}}
\caption{Comparison of (a) training and (b) prediction time with respect to the knowledge representation technique.}
\label{fig:time_training}
\end{figure}

Comparision of the training and prediction time taken by the knowledge structures is provided in Fig. \ref{fig:time_training} (a) and (b). 

We have compared our framework with four existing methods (\cite{shafizadeh2015performance}, \cite{lin2011predictive}, \cite{feng2016modeling} and \cite{kamusoko2015simulating}) in terms of $FoM, PA, UA$ and $OA$. Some of the distance based features which have been used in these works are distance to roads, built-up, river, urban planning area, central business district, railway, wetlands, forests, agricultural lands etc. Some of the non-distance based factors are digital elevation maps, slope, population density, land/crop density etc. During experimentation, we have manually generated each of these feature maps to model and compare the results of these works with our \textit{End-to-End} framework.

\subsection{Discussion}\label{section:discussion}

\begin{figure*}
\centering
\subfloat[][1991-2001]{\includegraphics[scale=0.5]{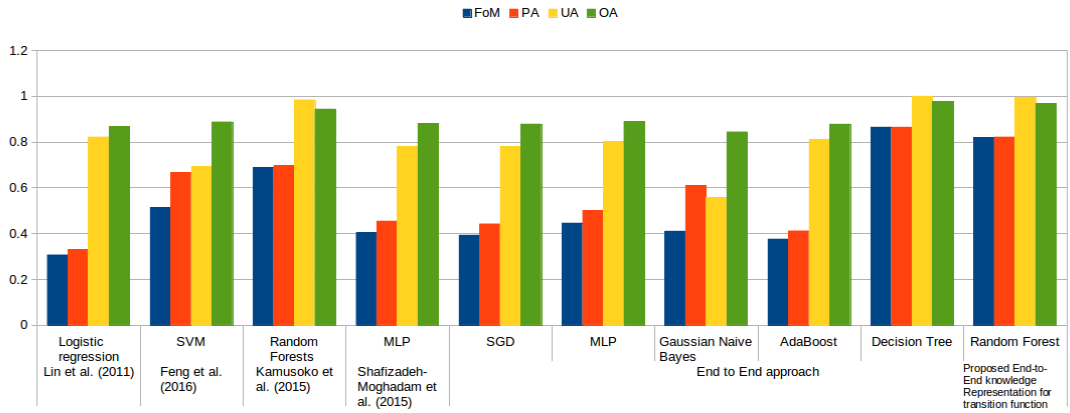}}

\subfloat[][2001-2011]{\includegraphics[scale=0.5]{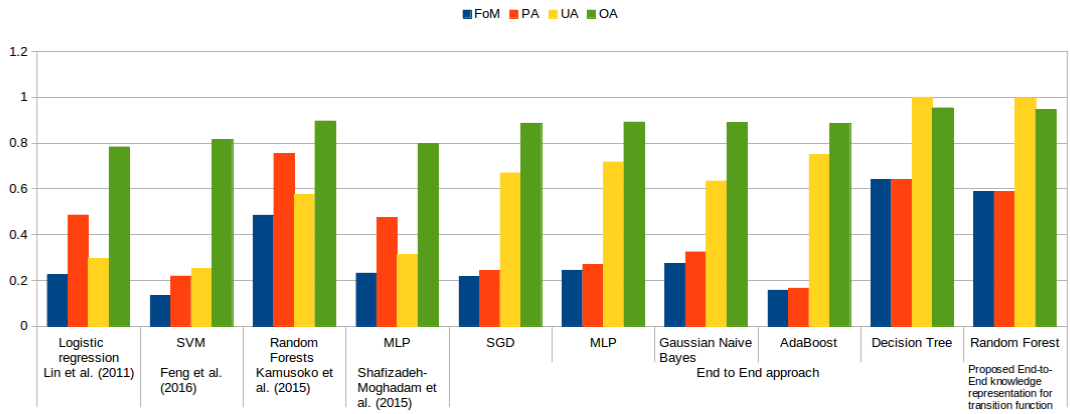}}
\caption{Performance comparison of our approach with existing methods for years (a) $1991-2001$ (b) $2001-2011$}
\label{fig:comparison_trset}
\end{figure*}

We defend our hypothesis regarding \textit{End-to-End} learning for prediction of urban growth with the results of our experiments. The comparison of our framework with existing frameworks based on the four parameters ($FoM, PA, UA, OA$) as in Fig. \ref{fig:comparison_trset} (a) and (b) reveals that \textit{End-to-End} learning performs significantly better than the existing learning based methods developed for urban growth prediction. We argue based on the results that this is possible due to the superior representation and robustness of encoded representations combined with an ensemble of decision trees. An approximate summary of enhancements provided by our \textit{End-to-End} framework on the dataset is given as follows.

\begin{itemize}
\item $18\%$ improvement on Figure of Merit $FoM$.

\item $17\%$ improvement on Producer's accuracy $PA$.

\item $2\%$ improvement on User's Accuracy $UA$.

\item $3\%$ improvement of Overall Accuracy $OA$.
\end{itemize}

The cross validation accuracies and the training, as well as prediction time is given in Table \ref{table:crossval}. It is evident from the table that both Decision trees and Random Forests (ensemble of decision trees) provide optimal results in terms of cross validation accuracy, training time, prediction time, $FoM, PA, UA$ and $OA$ as compared to other prediction models. However, Decision tree has a problem of overfitting which can be fixed by using a Random Forest. Hence we conclude that Random Forest (ensemble of decision trees) is an optimal choice for knowledge representation for our proposed \textit{End-to-End} framework.

\begin{figure}
\centering
\subfloat[][1991-2001]{\includegraphics[scale=0.5]{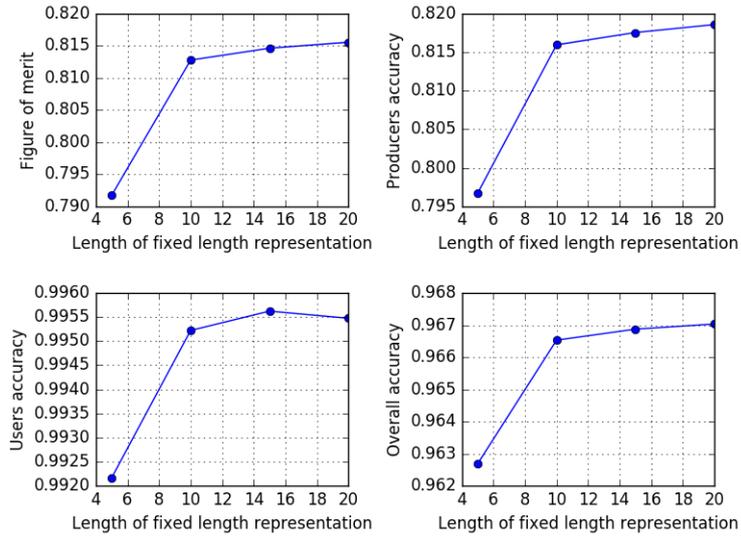}}

\subfloat[][2001-2011]{\includegraphics[scale=0.5]{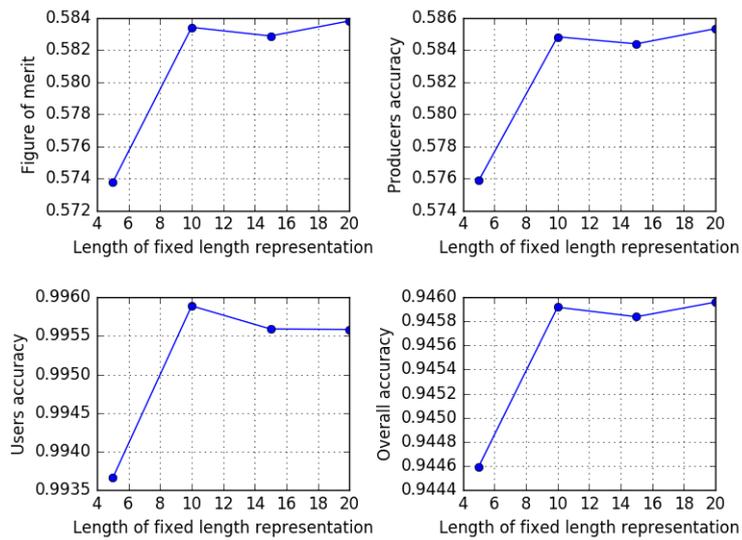}}
\caption{Performance metrics with respect to the encoding size.}
\label{fig:performance_encodingsize}
\end{figure}

Figure \ref{fig:performance_encodingsize} depicts the change of performance metrics with respect to the size of the encoding. It shows a sharp rise in the begining followed by saturation in the performance metrics as the encoding size is increased. This implies that with increase in encoding size, information is more precisely encoded by the autoencoder and hence performance of the simulation improves. However, the tradeoff is if encoding size is increased arbitrarily then time required to train the autoencoder increases more than performance metrics. Furthermore, increasing the encoding size increases size of the feature vector in the \textit{data matrix}, which brings in the \textit{curse of dimensionality} issue\footnote{When dimensionality of a feature vector is increased without increase in size of the data, then data tends to become sparse. This problem is referred to as \textit{curse of dimensionality}}. In this case, we believe that saturation is caused by increase in dimensions of the feature vector in the \textit{data matrix}.

From the comparison in Fig. \ref{fig:comparison_trset} (a) and (b), we have observed that the Overall Accuracy (OA) and the User's Accuracy (UA) metric is much higher than the Figure of merit (FoM) and Producer's accuracy (PA) for certain existing methods. This is indeed the case that has also been reflected in results shown by experiments in some of the other works, for instance, \cite{shafizadeh2015performance} have claimed $0.435, 0.6062$ and $0.8513$ as FoM, PA and OA using MLP for the region of Mumbai. One of the reasons behind this peculiarity is due to the imbalance in the datasets problem that we discussed earlier in the paper (section III A). Since Figure of Merit and Producer's accuracy provides the performance of the model in terms of transitions and the fraction of transition pixels is low, therefore a training algorithm might not have learned them correctly. Furthermore, we can also see that existing models have comparable Overall Accuracy (OA) of about $90\%$. This can be due to the fact that only $10\%$ of the pixels fall in the transition class and the simple default strategy to give maximum accuracy is to give $90\%$ \cite{chawla2002smote}. In the \textit{End-to-End} approach, we have seen that the $3$ metrics $FoM,PA$ and $UA$ are comparable, which indicates that the imbalance in datasets have been handled. 

The user's accuracy metric is high for certain models because of the fact that there is only one direction of change i.e. \textit{non built-up} to \textit{built-up}. Therefore, predicted pixels in the transition category can be only one category and the model that predicted a higher number of pixels in the transition category have higher user's accuracy.

\begin{table*}
\centering
\resizebox{\textwidth}{!}{\begin{tabular}{|c|c|c|c|}
\hline
{\bf Classifier} & {\bf Cross-Validation (Mean +/- variance)} & {\bf Training Time (in seconds)} & {\bf Prediction Time (in seconds)}\\
\hline
MLP \cite{shafizadeh2015performance} & $0.899112 (+/- 0.041553)$ & $244.40$ & $0.46$\\
\hline
Logistic Regression \cite{lin2011predictive} & $0.891636 (+/- 0.039668)$ & $3.79$ & $0.03$\\
\hline
SVM \cite{feng2016modeling} & $ 0.628495 (+/- 0.111121)$ & $58284.61$ & $2184.21$\\
\hline
Random Forests \cite{kamusoko2015simulating} & $ 0.851369 (+/- 0.056350)$ & $3.13$ & $0.09$\\
\hline
\textit{End-to-End} approach (MLP single layer) & $0.911805 (+/- 0.029091)$ & $233.98$ & $3.7$\\
\hline
\textit{End-to-End} approach (SGD) & $0.902009 (+/- 0.025873)$ & $4.84$ & $0.14$\\
\hline
\textit{End-to-End} approach (Naive Bayes) & $0.853874 (+/- 0.120939)$ & $0.87$ & $1.58$\\
\hline
\textit{End-to-End} approach (KNN) & $0.826471 (+/- 0.051649)$ & $5603.82$ & $22524.38$\\
\hline 
\textit{End-to-End} approach (AdaBoost) & $0.850093 (+/- 0.016626)$ & $138.40$ & $12.59$\\	
\hline
\textit{End-to-End} approach (Decision Tree) & $0.849341 (+/- 0.049533)$ & $31.44$ & $0.36$\\
\hline
{\bf \textit{End-to-End} approach (Random Forests)} & {\bf 0.900792 (+/- 0.043120)} & {\bf 53.06} & {\bf 2.9}\\
\hline
\end{tabular}}
\caption{Performance of other classifiers in comparison to Decision tree and Random Forests}
\label{table:crossval}
\end{table*}

\begin{table*}
\centering
\resizebox{\textwidth}{!}{\begin{tabular}{|c|c|}
\hline
{\bf Model} & {\bf Remarks}\\
\hline
MLP (\cite{shafizadeh2015performance}) & \specialcell{High Training time. Poor FoM, PA and UA than \\Proposed Method. Uses feature engineering.}\\
\hline
Logistic Regression (\cite{lin2011predictive}) & \specialcell{Poor FoM, PA and UA than Proposed Method.\\ Uses feature engineering.}\\
\hline
SVM (\cite{feng2016modeling}) & \specialcell{High training times. Poor FoM, PA and UA than \\Proposed Method. Uses feature engineering.}\\
\hline
Random Forests (\cite{kamusoko2015simulating}) & \specialcell{Poor FoM, PA and UA than Proposed Method. \\Uses feature engineering.}\\
\hline
\textit{End-to-End} approach (MLP single layer) & High Training time. Poor FoM, PA and UA than Proposed Method.\\
\hline
\textit{End-to-End} approach (SGD) & Poor FoM, PA and UA than Proposed Method.\\
\hline
\textit{End-to-End} approach (Naive Bayes) & Poor FoM, PA and UA than Proposed Method.\\
\hline
\textit{End-to-End} approach (Decision Tree) & Optimal but has Problem of overfitting.\\
\hline
{\bf \textit{End-to-End} approach (Random Forests)} & {\bf Optimal}\\
\hline
\end{tabular}}
\caption{Limitations of the other models to the proposed model.}
\label{table:limit}
\end{table*}

\subsection{Future Urban Growth Prediction}

\begin{figure*}
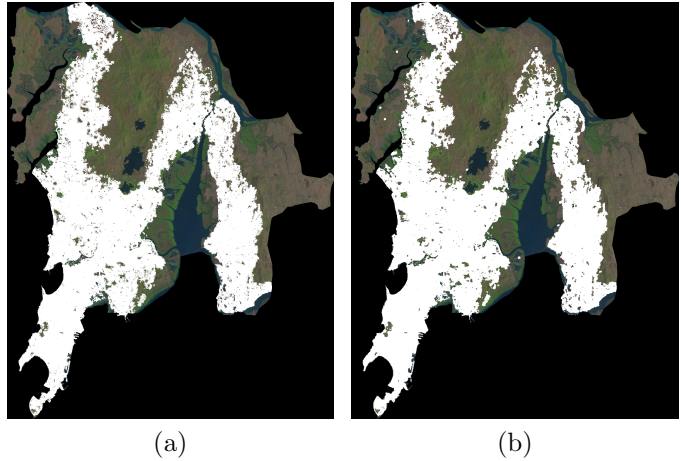

\centering
\subfloat[][]{\includegraphics[scale=0.10]{\locimage{mumbai_DT_2051}}}\hspace{0.1cm}
\subfloat[][]{\includegraphics[scale=0.10]{\locimage{mumbai_RF_2051}}}
\caption{Simulated urban growth of 2051 on mumbai region with knowledge structure as (a) Decision Tree and (b) Random Forests}
\label{fig:future_sim}
\end{figure*}

Figures \ref{fig:future_sim}(a) and \ref{fig:future_sim}(b) show future urban growth prediction for year $2051$ starting from $1991$ using our \textit{End-to-End} framework and knowledge structure as Decision Tree as well as Random Forest respectively. The raster over which the built-up is displayed is the year $1991$ natural color image of Mumbai. The white pixels represent the \textit{built-up} regions. It can be seen that the framework in case of Decision tree does not encroach upon water bodies and swamps, whereas in case of Random Forest, few encroachments are present. It may be noted that despite providing no explicit region information like water bodies, swamps, forests etc to the framework during training, the framework has been able to capture them. Hence, we can say that the rule based framework automatically divides the regions in the satellite image in some way to determine in which places growth can happen. This knowledge is currently encoded in the decision trees which can be extracted only if we discover the rules.

%% file: main.bbl
\newcommand{\noopsort}[1]{} \newcommand{\printfirst}[2]{#1}
  \newcommand{\singleletter}[1]{#1} \newcommand{\switchargs}[2]{#2#1}